\documentstyle[preprint,psfig,aps]{revtex}
\begin{document}                
\def\a{$\alpha$}
\def\be{\begin{equation}}
\def\ee{\end{equation}}
\def\ba{\begin{eqnarray}}
\def\ea{\end{eqnarray}}
\preprint{DOE/ER/40561-354-INT98-00-2}
\title{Bremsstrahlung in $\alpha$--Decay}
\author{Thomas Papenbrock and George F. Bertsch}
\address{Institute for Nuclear Theory, Department of Physics, 
University of Washington, Seattle, WA 98195, USA}
\maketitle
\begin{abstract}
We present the first fully quantum mechanical calculation of photon radiation 
accompanying charged particle decay from a barrier resonance. 
The soft--photon limit agrees with the classical results, 
but differences appear at next--to--leading--order.
Under the conditions of \a--decay of heavy nuclei,
the main contribution to the photon emission stems from Coulomb 
acceleration and may be computed analytically.
We find only a small contribution from the 
tunneling wave function under the barrier. 
\end{abstract}
Nuclear fission and $\alpha$--decay are interesting processes that involve 
both tunneling and the acceleration of charged particles in Coulomb fields.
This raises the question of whether the tunneling process affects the 
bremsstrahlung emission. A semiclassical theory with an affirmative conclusion
has been given by Dyakonov and Gornyi in ref.~\cite{Dyakonov}. 
In an experiment on the spontaneous fission of ${}^{\rm 252}$Cf, 
Luke~{\em et. al.}~\cite{Luke} found a null result and gave an upper limit
to the bremsstrahlung rate. 
In the case of $\alpha$--decay of heavy nuclei, two recent experiments
by D'Arrigo~{\em et. al.}~\cite{DArrigo}, and 
Kasagi~{\em et. al.}~\cite{Kasagi} detected accompanying photon radiation. 
The latter authors claimed to observe interference effects with a tunneling 
contribution to the bremsstrahlung, interpreting their results in the framework
of ref.~\cite{Dyakonov}. This gives urgency to carry out a full quantum 
mechanical calculation of the bremsstrahlung. We describe here a calculation 
with the following assumptions: a single--particle barrier model to 
describe the $\alpha$--nucleus wave function;
initial state treated as a Gamov state with a complex energy;
perturbative coupling of the photon to the current, taken in the dipole
approximation.  

It is convenient to use the acceleration form of the dipole operator,
\be
\langle f|{\vec p}|i\rangle 
={1\over E_\gamma} \langle f|[H,{\vec p}]|i\rangle 
={i\hbar\over E_\gamma} \langle f|\nabla V|i\rangle ,
\ee
where $H$ is the \a--particle Hamiltonian with potential $V$ and
$E_\gamma$ is the transition energy between initial and final state 
$|i\rangle$ and $|f\rangle$, respectively. 
The perturbative expression for dipole photon emission during
decay may be obtained 
from Fermi's golden rule and is given by 
\be
\label{fermi}
{dP\over d E_\gamma} = \frac{4 Z_{\rm \tiny eff}^2e^2}{3 m^2 c^3}
  \left|\langle\Phi_f\left|\partial_r V\right|\Phi_i\rangle\right|^2
  {1\over E_\gamma}.
\ee
Here ${dP/dE_\gamma}$ is the branching ratio to decay with a photon
emission, differential in the photon energy $E_\gamma$.  
The wave functions $\Phi_i(r)$ and 
$\Phi_f(r)$ are the radial wave functions of the initial 
and final state of the $\alpha$--particle, respectively, with normalizations
specified below.  The effective charge $Z_{\rm \tiny eff}$ for 
dipole transitions is given by $Z_{\rm \tiny eff} =((A-4)z-4(Z-2)/A$
where $z=2$ is the charge of the \a--particle and $Z, A$ are the
charge and mass number of the decaying nucleus, respectively.

We take the potential in the single--particle Hamiltonian as the Coulomb
outside a radius $r_0$ and a constant inside,
\be
\label{pot}
V(r) = \frac{Zze^2}{r}\Theta(r-r_0) - V_0\Theta(r_0-r). 
\ee
We will see later that the results are quite insensitive to the choice of 
parameters $V_0$ and $r_0$, provided the decay properties are reproduced.
The initial state $\Phi_i$ is a resonant state of zero angular 
momentum normalized to a unit outgoing flux of particles. Its radial 
wave function is given in terms of $F_0$ and $G_0$ Coulomb wave functions by
$\left({m \over\hbar k}\right)^{1\over 2}     
 \left[G_0(\eta,kr)+iF_0(\eta,kr)\right]/r$ outside $r_0$ 
and is proportional to the $j_0(\kappa r)$ spherical Bessel function inside.
The Sommerfeld parameter $\eta$ is given by $
\eta = \frac{zZe^2m}{\hbar^2k}$;
it is much larger than one for heavy nuclei. The wave numbers $k$ and
$\kappa$ satisfy
$k=\hbar^{-1}\sqrt{2mE_\alpha}$
and
$\kappa=\hbar^{-1}\sqrt{2m(E_\alpha+V_0)}$
where $E_\alpha$ is the \a--decay energy.
Matching of the wave functions at $r=r_0$ yields
the amplitude of the inner wave function as well as 
the (complex) energy of the resonant state.  

The parameters $r_0$ and $V_0$ of our nuclear potential 
(\ref{pot}) are fixed to reproduce the empirical decay energy
$E_\alpha$ and mean life $\tau$ of the decay.  
The mean life time depends on the parameters through the equation
\be
\label{width}
\frac{2E_\alpha\tau}{\hbar} \approx 
\frac{kr_0}{2}\frac{G_0^2(\eta,kr_0)}{\sin^2\kappa r_0}
\left(1 - \frac{\sin 2\kappa r_0}{2\kappa r_0}\right) 
+ \int\limits_{kr_0}^{2\eta} d\rho\,G_0^2(\eta,\rho).
\ee
In eq.~(\ref{width}) and also in the wave function matching, we neglect 
terms with $F_0$ and $F'_0$ which are of order 
$O(\Delta)\ll 1$ compared to $G_0$ and $G'_0$. 
Here $\Delta={\hbar\over\tau E_\alpha}$ is a small parameter, and the primes
denote derivatives with respect to $kr$. 

As is well known, there are multiple solution sets $(r_0,V_0)$ for a given 
decay energy and mean life, distinguished by the number of nodes of the 
inner wave function~\cite{Bu90}. Typical solution sets for the nuclei of 
interest are shown in Table~\ref{tab1}. Our simple model gives reasonable
radii close to or slightly larger than the nuclear radius for $V_0$ in the
range 0 to 150 MeV~\cite{Bu90}. The results presented below do not depend
on a specific choice of a solution.  

The continuum final states $\Phi_f$ are normalized to give the
unit operator when integrated over energy, $\delta(r-r') =
\int d E \Phi_E(r)\Phi_E(r')$.  The radial wave function is a standing
wave having the form 
$\left(2m\over\pi\hbar^2 k'\right)^{1\over 2}
\left(G_1(\eta',k'r)\sin\alpha+F_1(\eta',k'r)\cos\alpha\right)/r$
outside $r_0$ and proportional to the $j_1(\kappa'r)$ spherical Bessel
function inside. Primed quantities are defined similar to the corresponding 
unprimed quantities, but with the energy diminished by the emission of a 
photon.
Matching the wave function at $r=r_0$ yields expressions for  
$\tan\alpha$ and the amplitude of the inner wave function.
The final state is off resonance for almost all final energies 
$(E_\alpha-E_\gamma)$, and thus 
$\tan\alpha \sim O(\Delta)\ll 1$, i.e. the final wave 
function does not penetrate the nucleus significantly
and is a true continuum wave function. Outside the Coulomb--barrier the wave 
function is very well approximated by the regular Coulomb wave function only.

The matrix element in eq. (\ref{fermi}) can now be written down.  It has
a delta function contribution at $r_0$ and an integral over the
derivative of the Coulomb field outside,
\ba
\label{matele}
&&\langle\Phi_f\left|\partial_r V\right|\Phi_i\rangle 
= \sqrt{2m^2\over\pi\hbar^3kk'}\Bigg\{\left(\frac{zZe^2}{r_0}+V_0\right)
\nonumber\\
&&\times\left[F_1(\eta',k'r_0)+G_1(\eta',k'r_0)\tan\alpha\right]
  G_0(\eta,kr_0) \nonumber\\
&&-zZe^2\int_{r_0}^\infty\!dr\,r^{-2}\Big\{
  \left[F_1(\eta',k'r)+G_1(\eta',k'r)\tan\alpha\right]
\nonumber\\
&& \times\left[G_0(\eta,kr)+iF_0(\eta,kr)\right]\Big\}\Bigg\}
+ O(\Delta).
\ea
We separate the expression (\ref{matele}) into real and imaginary 
contributions and consider the latter first.

We may neglect the contribution of the term 
$F_0(\eta,kr)G_1(\eta',k'r)\tan\alpha$ to the integral since it is 
of order $O(\Delta)$. Thus, the imaginary part is an integral over 
two $F_j$ functions.  Therefore it contains those contributions to the 
bremsstrahlung that stem from the classical acceleration in the Coulomb 
field. It can be treated analytically as follows.  
We first extend the lower limit of the integral to zero, which only introduces 
an error of the order $O(\Delta)$. The resulting 
integral may be expressed in terms of hypergeometric functions as
\cite{Biedenharn,Alder} 
\ba
\label{coulint}
&&\int_{0}^\infty\frac{dr}{r^2}F_1(\eta',k'r)F_0(\eta,kr) 
\nonumber\\&=& 
kk'\left\{k'\,|1+i\eta'|\,M_0\,-\,k\,|1+i\eta|\,M_1\right\}
\ea
where
\ba
M_j&=&
\left(\frac{\xi}{\eta+\eta'}\right)^{i(\eta+\eta')}
\frac{|\Gamma(j+1+i\eta')|\,|\Gamma(j+1+i\eta)|}{(k-k')^2(2j+1)!}
\nonumber\\
&\times& e^{-\frac{\pi}{2}\xi}\,\,
\left(\frac{\eta'\eta}{\xi^2}\right)^j\,\,{}_2\!
F_1\left(j+1-i\eta,j+1-i\eta',2j+2;-\frac{\eta'\eta}{\xi^2}\right).
\ea
Here ${}_2\!F_1$ denotes the hypergeometric function and we have defined 
\cite{comment}
\be
\label{xi}
\xi = \eta'-\eta.
\ee
In the limit of vanishing photon energy the imaginary part of the matrix
element (\ref{matele}) may be computed directly, using \cite{Arnoldus}.
\be
\label{Imzero}
\lim_{k'\to k} {\rm Im}\langle\Phi_f\left|\partial_r V\right|\Phi_i\rangle
= -\sqrt{\frac{mE_\alpha}{\pi\hbar}}\frac{\eta}{\sqrt{1+\eta^2}}.
\ee
The real part of the matrix element (\ref{matele}) is a sum of two terms
which, in contrast to the imaginary part, involve contributions from the 
irregular Coulomb wave functions $G_j$. Thus, it describes those 
contributions to the bremsstrahlung that are associated with tunneling.
In the limit of vanishing photon energy, this amplitude reduces 
to ~\cite{Arnoldus,comment2} 
\be
\label{Rezero}
\lim_{k'\to k} {\rm Re}\langle\Phi_f\left|\partial_r V\right|\Phi_i\rangle
= \sqrt{\frac{mE_\alpha}{\pi\hbar}}\frac{1}{\sqrt{1+\eta^2}}.
\ee
Notice that the dependence on the inner barrier parameters has disappeared.
A comparison with the imaginary part (\ref{Imzero}) shows that the real 
part (\ref{Rezero}) is suppressed by a factor $\eta$. For nonzero 
photon energy we have to treat the real part of the matrix element 
(\ref{matele}) numerically. However, the numerical evaluation shows 
that the real part still is suppressed in comparison to the imaginary part.
This implies that only a smaller fraction of bremsstrahlung is 
emitted during tunneling. Note also that the contributions associated 
with classical acceleration and tunneling do not interfere since they 
differ in phase by $i$.

We will now make the connection to semiclassical and classical
limits. For heavy nuclei, the
Sommerfeld parameters $\eta$ are large and the Coulomb wave
functions $F_j$ may be approximated by their WKB--wave functions
\be
F^{\mbox{\em\tiny WKB}}_j(\eta,kr)=\left(k^2/f(r)\right)^{1\over 4}\sin\phi,
\ee
with
\ba
f(r)&=&k^2-2k\eta/r - j(j+1)/r^2\quad\mbox{and}\\
\phi&=&\frac{\pi}{4}+\int_{2\eta/k}^{r}dr'\,[f(r')]^{\frac{1}{2}}.
\ea
In leading order in  $\eta, \eta'$ one finds \cite{Alder}
\ba
\label{semicl}
&&\int\limits_{2\eta}^\infty\frac{dr}{r^2}
F^{\mbox{\em\tiny WKB}}_1(\eta',k'r)F^{\mbox{\em\tiny WKB}}_0(\eta,kr) 
\nonumber\\
&\approx& -\frac{kk'}{k+k'}\frac{\xi}{\bar\eta} e^{-{\pi\over 2}\xi}
\left[K'_{i\xi}(\xi\epsilon)+
\frac{(\epsilon^2-1)^{1\over 2}}{\epsilon}K_{i\xi}(\xi\epsilon)\right],
\ea
where $\epsilon = {(\bar\eta^2 + 3/4)^{1\over 2}\over\bar\eta}$ and 
$\bar\eta = (\eta'+\eta)/2$.
$K_{\nu}(z)$ denotes the modified Bessel function and
$K'_{\nu}$ its derivative with respect to the argument. 

A comparison of the semiclassically evaluated integral 
(\ref{semicl}) with the quantum mechanical result (\ref{coulint}) shows
that they deviate from each other by less than one percent for photon energies
$E_\gamma$ up to 1 MeV. We recall that the semiclassical computation 
neglects any contributions from the wave functions at radii smaller
than the classical turning point, i.e. any contribution from the tunneling.
This clearly justifies the attribution of tunneling to the real part of the
matrix element, (\ref{matele}), alone. 

Next we consider the classical and the soft photon limit. 
The classical formula valid at all frequencies can be derived from 
\cite{Ja62}
\be
\label{bremscl}
{dP\over d E_\gamma} = 
\frac{2\alpha Z^2_{\rm \tiny eff}}{3\pi}\frac{|I(\omega)|^2}{E_\gamma}
\ee
with $I$ the Fourier transform of the time--dependent acceleration,
\be
I(\omega)=c^{-1}\int_0^\infty dt\,\frac{dv}{dt}\exp(i\omega t).
\ee
This integral can be expressed in terms of the dimensionless parameter
\be
\label{zeta}
\zeta=\eta\frac{\hbar\omega}{E_\alpha}
\ee
as
\be
\label{Iom}
I(\omega)=\sqrt{\frac{2E_\alpha}{mc^2}}\int\limits_0^1 dz\,
\exp\left[i\zeta\left({z\over 1-z^2}+{\rm artanh}\,z\right)\right].
\ee
In the limit of small photon energy we find 
\be 
\label{zeroerg}
{dP\over d E_\gamma} = 
\frac{4\alpha Z^2_{\rm \tiny eff}}{3\pi}\frac{E_\alpha}{mc^2} E_\gamma^{-1}.
\ee
Because this only depends on the asymptotic motion of the particles,
the quantum result must coincide. Inserting the results (\ref{Imzero}), 
(\ref{Rezero}) in eq. (\ref{fermi}) indeed yields the classical result, 
eq. (\ref{zeroerg}). 

\begin{figure}
  \begin{center}
    \leavevmode
    \parbox{0.9\textwidth}
           {\psfig{file=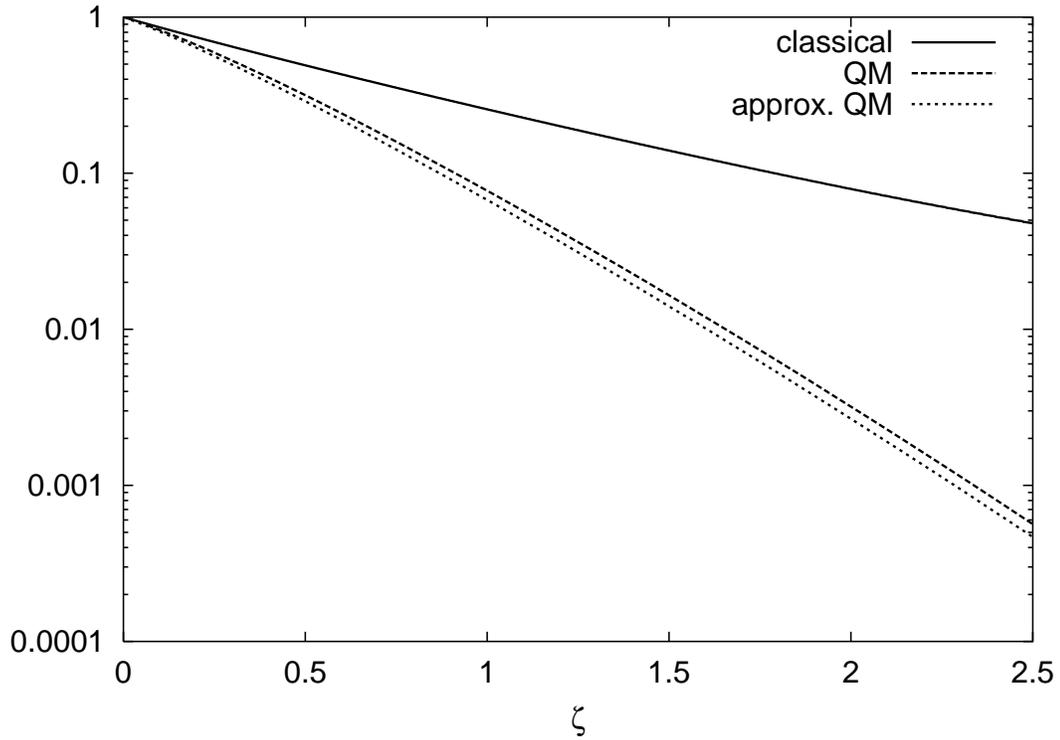,width=0.9\textwidth,angle=270}}
  \end{center}
\protect\caption{Comparison of classical and quantum mechanical photon
emission probability in \a--decay of ${}^{210}{\rm Po}$. 
Curves show the probabilities
normalized to the low--energy formula, eq. (\ref{zeroerg}), as a
function of the scaled photon frequency $\zeta$ defined in eq. (\ref{zeta}).
The classical probability, eq. (\ref{bremscl}), is shown as the solid line.  
The quantum probabilities (nearly exponentially falling lines) are shown 
for the full quantum mechanical treatment (dashed) and for the approximation 
that neglects contribution from tunneling (dotted). $\zeta=1$ corresponds to 
$E_\gamma\approx 0.24$MeV.}
\label{fig2}
\end{figure}

More interesting is to examine the next--to--leading $E_\gamma$--dependence
and compare the quantum and classical behavior. It turns out that the 
classical  parameter $\zeta$ in eq. (\ref{zeta}) is essentially the same
as the quantum small parameter $\xi$ defined in eq.~(\ref{xi}) 
($\zeta=2\xi+O(\xi^2)$). This parameter may also be identified with the
product of the photon frequency and the barrier tunneling time. The 
$\zeta$--dependence of the classical and quantum calculations are compared
in Fig.~\ref{fig2}. The solid line shows the classical 
prediction (\ref{bremscl}). The dashed and dotted lines show the 
quantum result with and without tunneling contributions, respectively.
We see that the tunneling contributions remains small even at a finite photon
energy.

One might have expected that the classical curve would be tangent to the 
quantum at $\omega=0$: in scattering bremsstrahlung is determined by 
on--shell amplitudes to next--to--leading order \cite{Low58}. We find that
the two curves are indeed very close in the neighborhood $\omega=0$, but
the slopes are not identical. For large photon energies the classical 
result overestimates the photon emission rate considerably since the classical 
formula (\ref{bremscl}) neglects any energy loss of the escaping 
$\alpha$--particle. This point has been discussed in the framework of photon 
emission in spontaneous fission by Luke {\em et. al.} \cite{Luke}, and 
earlier in the framework of Coulomb excitation by 
Alder~{\em et. al.}~\cite{Alder}.   

The quantum mechanical results for ${}^{214}$Po and ${}^{226}$Ra
are practically identical to those for ${}^{210}$Po when plotted as in 
Fig.~\ref{fig2}, normalized to the $\omega=0$ rate~(\ref{zeroerg}) and 
plotted as a function of $\zeta$. Since $\zeta$ is inversely proportional
to the decay energy $E_\alpha$, the rates are higher for higher decay
energies. Thus for ${}^{214}$Po decay, with an \a--decay energy of 7.7 MeV,
the predicted rate for $E_\gamma=0.6$ MeV is 65 times higher than for 
${}^{210}$Po.

\begin{figure}
  \begin{center}
    \leavevmode
    \parbox{0.9\textwidth}
           {\psfig{file=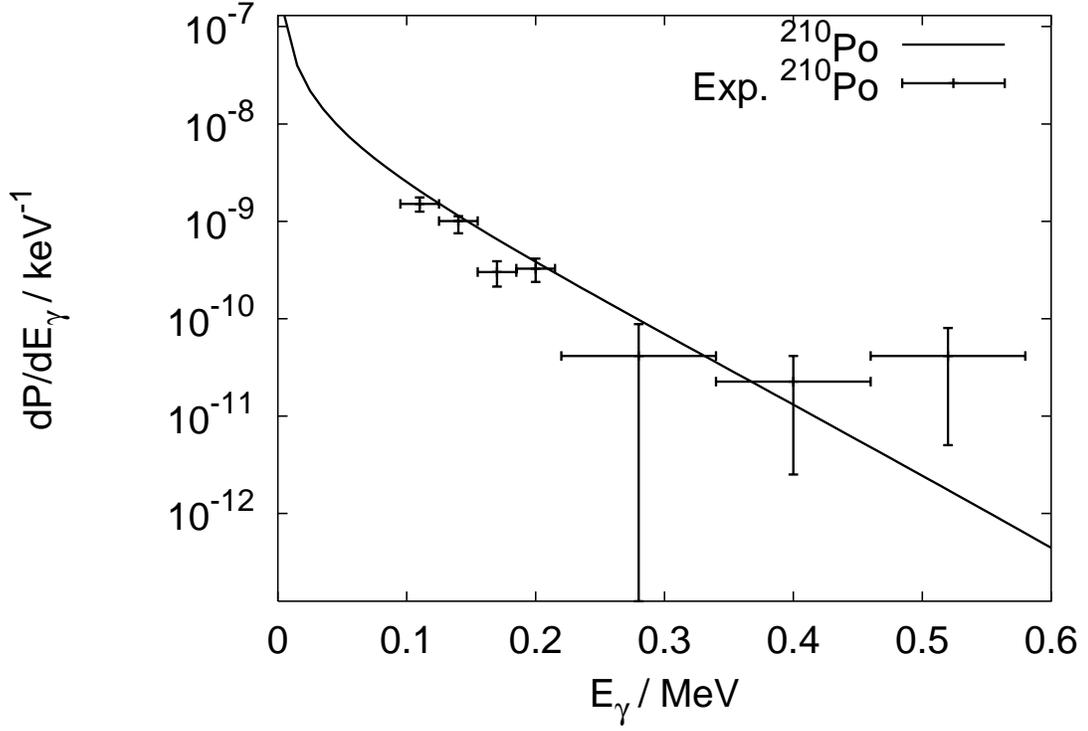,width=0.9\textwidth,angle=270}}
  \end{center}
\protect\caption{Photon emission probability comparing
the quantum calculation with experiment.
Logarithmic plot of $\frac{dP}{dE_\gamma}$ as a 
function of photon energy for ${}^{210}{\rm Po}$ (full line).  
The experimental data for ${}^{210}{\rm Po}$ (datapoints with errorbars)
are taken from Ref.~\protect\cite{Kasagi}.}
\label{fig1}
\end{figure}

Finally, we compare the results obtained in this work with experiment. 
In the case of ${}^{\rm 210}$Po, our result displayed in 
Figure~\ref{fig1} is consistent with the experimental result obtained by 
Kasagi~{\em et. al.}~\cite{Kasagi} suggesting that no interference resulting 
from photon emission during
tunneling is needed for an explanation of the experiment.  
In the case of ${}^{\rm 214}$Po and ${}^{\rm 226}$Ra, 
D'Arrigo~{\em et. al.}~\cite{DArrigo} reported photon emission rates that are
larger than expected from the classical formula (\ref{bremscl}). Thus, their
results are also more than one order of magnitude larger than our 
quantum mechanically computed rate.
We cannot trace the origin of this difference.

In summary,
we have used Fermi's golden rule to compute the emission of bremsstrahlung
in $\alpha$--decay of heavy nuclei. The dominant contribution to the photon
emission rate stems from classical acceleration and is given in closed form. 
Only a smaller fraction of bremsstrahlung is emitted during tunneling. 
This finding is consistent with experimental data on ${}^{\rm 210}$Po. 

We thank R. Vandenbosch, A. Bulgac and N. Takigawa for discussions. 
This work was supported by the Department of Energy on 
contract No. DE-FG-06-90ER-40561.

\begin{table}
\begin{tabular}{|c||c|c|c|}
                 & ${}^{214}{\rm Po}$ & ${}^{210}{\rm Po}$ & ${}^{226}{\rm Ra}$ \\\hline\hline
 $E/{\rm MeV}$   &   7.7          &     5.3        &    4.8         \\\hline
 $\Delta=\hbar/\tau E_\alpha$ & $3.7\cdot 10^{-19}$ & $7.3\cdot 10^{-30}$ & $1.9\cdot 10^{-33}$\\ \hline
 $r_0/{\rm fm}$  & 9.19          &    8.76        &    9.75        \\ \hline
 $V_0/{\rm MeV}$ & 12.1          &    16.7        &   12.9       \\ 
\end{tabular}
\protect\caption{Parameters of the nuclear potential.
The table lists the parameters $r_0$ and $V_0$ of the nuclear 
potential (\protect\ref{pot}) that were used in the present calculations.
For each listed nuclei, the presented parameters reproduce the experimentally 
known values of the $\alpha$--particle energy $E_\alpha$ and the mean life 
time $\tau$.} 
\label{tab1}
\end{table}

\begin{references}
\bibitem{comment}For small photon energies, 
$x_0=\left|{\eta'\eta\over\xi^2}\right| > 1$, and
the hypergeometric function is conveniently evaluated using
the identity  \cite{GR} relating it to $_2F_1(...,1/x_0)$; the latter may be
expanded in a power series in $1/x_0$.
\bibitem{comment2} We use the matching conditions, the recursion relations
of the Coulomb wave functions and spherical Bessel functions, and as their
result, 
${j'_1(\kappa r_0)\over j_1(\kappa r_0)}=
-\frac{G_0(kr_0)\kappa r_0}{G'_0(kr_0)kr_0-G_0(kr_0)}-{2\over\kappa r_0}$.
\end{references}
\end{document}